\documentclass[aps,pra,twocolumn,groupedaddress,showpacs,showkeys]{revtex4}

\usepackage{graphicx}
\begin{document}

\newcommand{\be}{\begin{equation}}
\newcommand{\ee}{\end{equation}}

\title{Comment on ''Quantitative wave-particle duality in multibeam interferometers"}

\author{G. Bimonte and R. Musto}
\email[Bimonte@na.infn.it, Musto@na.infn.it]
\affiliation{Dipartimento di Scienze Fisiche, Universit\`{a} di
Napoli Federico II, Complesso Universitario MSA, Via Cintia
I-80126 Napoli, Italy; INFN, Sezione di Napoli, Napoli, ITALY.}

\date{\today}

\begin{abstract}
In a recent paper [Phys. Rev. {\bf A64}, 042113 (2001)] S. D\"urr
proposed an interesting multibeam generalization of the
quantitative formulation of interferometric wave-particle duality,
discovered by Englert for two-beam interferometers. The proposed
generalization is an inequality that relates a generalized measure
of the fringe visibility, to certain measures of the maximum
amount of which-way knowledge that can be stored in a which-way
detector.   We construct an explicit example where, with three
beams in a pure state, the scheme proposed by D\"{u}rr leads to
the possibility of an ideal which-way detector, that can achieve a
better path-discrimination, at the same time as a better fringe
visibility. In our opinion, this seems to be in contrast with the
intuitive idea of complementarity, as it is implemented in the
two-beams case, where an increase in path discrimination always
implies a decrease of fringe visibility, if the beams and the
detector are in pure states.
\end{abstract}

\pacs{03.65.Ta, 03.65.Ud}
\keywords{interferometric, duality, multibeam}

\maketitle


\section{Introduction}
As it is well known, Bohr's {\it Principle of Complementarity},
and the subsequent debate  on the possibility of detecting, as
proposed by Einstein, "which-way" individual quantum systems
("quantons", for short) are taking, in double-slit interference
experiments,   helped to shape the basic concepts of Quantum
Mechanics. However, this early discussion on the duality between
fringe visibility and which-way information, as it is called
today, was essentially semiclassical in nature. The history of the
attempts of formulating such duality, for the two beams case,
within the full framework of Quantum Mechanics, has been quite
long, perhaps surprisingly long, and has found, it seems fair to
say, a satisfactory conclusion in 1996 in a paper by Englert
\cite{englert}. Following a suggestion present in the pioneering
work of Wootters and Zurek \cite{zurek}, Englert was able to
establish  a complementarity relationship between the {\it
distinguishability} ${\cal D}$, that gives a quantitative estimate
of the ways, and the {\it visibility} ${\cal V}$, that measures
the quality of the interference fringes: \be
  {\cal D}^2 +{\cal V}^2 \leq 1\;.\label{D+V}
\ee An important feature of   Eq.(\ref{D+V}) is that it becomes an
{\it equality} when the beams and the detector are prepared in a
pure state; when this is the case, Eq.(\ref{D+V}) implies that a
larger visibility is necessarily accompanied by a smaller path
distinguishability.

It is interesting to explore if an analogous form of
interferometric duality can be formulated for more than two beams
of interfering quantons.    An important step toward the
understanding of this question has been made by D\"{u}rr
\cite{durr}: he argued that an appropriate multibeam
generalization of the usual concept of fringe visibility ${\cal
V}$, is provided by the (properly normalized) rms spread $V$ of
the fringes intensity from its mean value (Eq.(1.10) of
Ref.\cite{durr}). By a corresponding generalization of the concept
of path predictability ${\cal P}$, provided by the quantity $P$
defined in Eq.(1.16) of Ref.\cite{durr}, D\"urr was able to derive
an inequality analogous to that found by Greenberger and Ya Sin
\cite{gree} for two beams: \be {P}^2 + {V}^2 \le 1\;.\label{PV}\ee
Similarly to Eq.(\ref{D+V}), the above inequality becomes an
equality if the beams are in a pure state, which ensures the
existence of a general see-saw relation between $V$ and $P$. Since
$V$ and $P$ undoubtedly  measure, respectively, wave-like and
particle-like attributes of the interfering quantons, we thus
think that Eq.(\ref{PV}) can be correctly interpreted as
expressing a form of wave-particle duality in the multibeam case.

However interesting, an inequality like Eq.(\ref{PV}) does not
convey yet the concept of wave-particle duality, as it is
involved, say, in the famous ideal experiment with two moving
slits, conceived by Einstein. Indeed, the quantity $P$ above does
not represent any real knowledge of the paths followed by
individual quantons, but only constitutes some measure of one's a
priori ability to predict them, based on unequal populations of
the beams.  The relevant schemes for a discussion of wave-particle
duality a' la Einstein, are those in which one actually tries to
obtain which-way knowledge, by placing detectors along the paths
of the quantons. In order to measure the amount of which-way
information, that can be obtained by measuring the detector's
observable $W$, after the passage of each quanton, D\"urr defines
the which-way knowledge $K(W)$ as a weighted average of the
generalized predictabilities $P$ of the sorted subensembles of
quantons, for which a certain result of the measurement is
obtained (Eqs.(2.3) and (2.4) of Ref.\cite{durr}) (Actually,
D\"urr introduces also an alternative measure $I_{KW}$ of the
which-way information, in Eq. (6.6) of Ref. \cite{durr}. For the
sake of simplicity, in this Comment, we will refer only to the
first one, and we address the interested reader to Ref.
\cite{bimonte}, where an extensive discussion of the problem is
given). Then, the multibeam analogue, $D$, of Englert's path
distinguishability ${\cal D}$ is defined as the maximum value of
$K(W)$, over the set of all detector's observables (Eq.(2.11) of
Ref.\cite{durr}).  By using this definition, D\"urr is able to
prove an inequality analogous to Englert's Eq.(\ref{D+V}): \be D^2
+ V^2 \le 1\;.\label{durr}\ee
\\ \indent This generalization of Eq.(\ref{D+V})
to the multibeam case, is an interesting relation, that can be
tested, in principle, by experiments. However, there exists a
difference between the two beams and the multibeam case. In fact,
differently from Eqs.(\ref{D+V}) and (\ref{PV}), the inequality
(\ref{durr}), cannot be saturated in general, even if the beams
and the detector are prepared in pure states (in Ref.
\cite{bimonte}, we actually prove that in the multibeam case the
above inequality can be saturated only if the visibility $V$ is
either equal to one or to zero). Therefore, one may conceive the
possibility of designing two which-way detectors $D_1$ and $D_2$,
such that $V_1
> V_2$, while, at the same time, $D_1 > D_2$.

It is the purpose of this Comment to show that this  possibility
actually occurs, as will be seen in next Section, by an explicit
example. In the final considerations that close this Comment, we
argue that such a behavior rises doubts on the possibility of
interpreting Eq.(\ref{durr}) as a statement of wave-particle
duality.

\section{A three-beam  example.}
\setcounter{equation}{0}

In this Section the problem announced in the previous Section is
presented  in an example with three beams of quantons in a pure
state. So, we consider a three beam interferometer with  equally
populated beams, described by the pure state: \be
\rho=\frac{1}{3}\sum_{i,j=1}^3    \;|\psi_i><\psi_j|\;. \ee If a
detector, initially prepared in some pure initial state
$|\chi_0>$, is placed along the trajectories followed by the
quantons, its interaction with the quantons will give rise to an
entangled state $\rho_{b \& d}$ of the form: \be \rho_{b \& d}=
\frac{1}{3}\sum_{i,j=1}^3|\chi_i><\chi_j| \otimes
|\psi_i><\psi_j|\;,\ee where $|\chi_i>$ are normalized, but not
necessarily orthogonal, detector's states. Suppose, for
simplicity, that the detector's Hilbert space ${\cal H}_D$ is
two-dimensional. In order to further specify the states
$|\chi_i>$, it is then convenient to use the Bloch
parametrization, to represent rays of ${\cal H}_D$ by unit
three-vectors, $\hat{n}=(n^x,n^y,n^z)$, via the map: \be
\frac{1+\hat n\, {\cdot} \,\vec{\sigma}}{2}\,
=|\chi><\chi|\;,\label{block} \ee where $\vec{\sigma}=(\sigma_x,
\sigma_y, \sigma_z)$ is any representation of the Pauli matrices
in ${\cal H}_D$. We shall denote by $|\hat{n}><\hat{n}|$ the ray
corresponding to the vector $\hat n$. We require that the
directions $\hat{n}_+,\hat{n}_-,\hat{n}_0$, associated with states
$|\chi_i>$, are coplanar, and such that $\hat{n}_+$ and
$\hat{n}_-$ both form an angle $\theta$ with $\hat{n}_0$. We
imagine that $\theta$ can be varied at will, by acting on the
detector. By properly choosing the orientation of the coordinate
axis, we can  make the vector ${\hat n}_0$ coincide with the $z$
axis, and the vectors ${\hat n}_{{\pm}}$ lie in the $xz$ plane,
such that: \be \hat{n}_0\equiv
(0,0,1)\;\;,\;\;\;\hat{n}_{{\pm}}=({\pm}
\sin\theta,\,0\,,\cos\theta)\;. \ee Upon using the well known
formula $|<{\chi_i}|{\chi_j}>|^2=(1+\hat{n}_i{\cdot}
\hat{n}_j)/{2}\;,$ into D\"urr's definition for the generalized
fringe visibility $V$, Eq. (1.12) of Ref. \cite{durr},
 one gets the following expression for $V$, as a function of $\theta$: \be V(\theta)=\sqrt{\frac{1}{6}\sum_i
\sum_{j \neq i}(1+\hat{n}_i {\cdot}\hat{n}_j) }=
\sqrt{\frac{1+\cos\theta+\cos^2 \theta}{3}}\;.\label{visex} \ee We
notice that the value of the visibility is equal to one, for
$\theta=0$, and gradually decreases when $\theta$ is increased,
until it reaches its minimum for  $\theta = 2 \,\pi/3$.
Afterwards, it starts increasing and keeps on increasing until
$\theta=\pi$ (see Figure).

The next step is to evaluate the generalized path
distinguishability $D$ as a function of $\theta$. This requires
that we determine the observable $W_{\rm opt}$ in ${\cal H}_D$
that maximizes the multibeam generalization of the which-way
knowledge $K(W)$. We briefly recall the definition of $K(W)$
proposed in Ref.\cite{durr}. Consider any detector's observable
$W$, and let $\Pi_l,\;(l=+,-)$ the projector onto the subspace of
${\cal H}_D$, relative to the eigenvalue $w_l$.  For any $W$, we
let $\hat m \equiv( \sin \beta \cos \gamma, \sin \beta \sin
\gamma, \cos \beta)$ the unique unit three-vector such that \be
\Pi_+=\frac{1+\hat m \cdot
\vec{\sigma}}{2}\;,\;\;\;\Pi_-=\frac{1-\hat m \cdot
\vec{\sigma}}{2}\;.\ee We let now $p_{i|l}$ the conditioned
probability to find a quanton in beam $i$, provided that the
measurement of $W$ on the which-way detector gave the outcome
$w_l$. According to Bayes' formula: \be p_{i|l}= \frac{\zeta_i\,
q_{i|l}}{p_{\,l}}\;, \ee where $q_{i|l}$ is the probability of
getting the outcome $w_l$, when the  quanton occupies with
certainty the beam $i$, while $\zeta_i$ are the populations of the
beams, and   $p_{\,l}$ is the total a-priori probability for
obtaining the result $w_l$, $p_{\,l}=\sum_i \zeta_i\, q_{i|l}$.
Recall that, in the above equation, we have to set $\zeta_i=1/3$,
because we are considering three equally populated beams.
According to Ref.\cite{durr}, the which-way knowledge $K(W)$
delivered by $W$, is  the weighted average $K(W)=\sum_l p_{\,l}
K_l$ of the partial predictabilities $K_l$ for the sorted
subensembles of quantons: \be
K_l=\sqrt{\frac{n}{n-1}\sum_i\left(p_{i|l}-\frac{1}{n}
\right)^2}\;.\ee Now,  using the well known formula, \be
q_{i|\pm}=<\chi_i|\Pi_{\pm}|\chi_i>=\frac{1\pm\hat{m}{\cdot}
\hat{n}_i}{2}\;,  \ee it is easy to verify that: \begin{widetext}
\be K^2=\frac{4}{9} \left[ \cos^2 \beta \,\sin^2\left(
\frac{\theta}{2}\right) + 3 \sin^2 \beta \cos^2 \gamma\,\cos^2
\left(\frac{\theta}{2}\right)
\right]\sin^2\left(\frac{\theta}{2}\right)\;. \ee
\end{widetext} For all values of $\theta$, the which-way
information is maximum if $\cos \gamma={\pm} 1$, i.e. if the
vector $\hat{m}_1$ lies in the same plane as the vectors
$\hat{n}_i$. As for the optimal value of $\beta$, it depends on
$\theta$. For $0 \le \theta < 2 \pi/3$, the best choice is $
\beta={\pm}\pi/2$,  and so for the optimal observable $W_{\rm
opt}$ we can take any operator such that: \be {\Pi}_{\pm}=
\frac{1\pm\sigma_x}{2}\;\;\;\;\;\;\;{\rm for}\;\;\; 0 \le \theta <
2 \pi/3\;,\label{pvmsm} \ee which delivers an amount of which-way
knowledge $D$ equal to:   \be \;\;\,D(\theta)=\,\frac{1}{\sqrt 3}
\,\sin \theta \;\;\;\;\;\;\;\;\;\;\;{\rm for}\;\;\;\;\;  0 \le
\theta <2/3\, \pi\;.\label{smal} \ee
\begin{figure}
\includegraphics{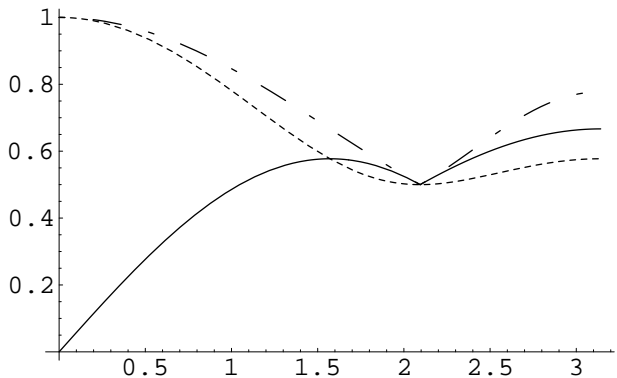}
\caption{\label{}Plots of the quantities $D$ (solid line), $V$
(dotted line), and  $D^2+V^2$ (dashed line), as functions of
$\theta$, in an ideal three-beam interference experiment.}
\end{figure}
For larger values of $\theta$,  the maximum information is reached
for $\beta=0$ and then the optimal operators are those for which:
\be {\Pi}_{\pm}= \frac{1\pm\sigma_z}{2}\;\;\;\;\;\;\;{\rm for}
\;\;\;2 \pi/3 < \theta \le \pi \;,\label{opt}\ee which deliver an
amount $D(\theta)$ of which-way information equal to:  \be
D(\theta)=\frac{2}{3} \sin^2 \left(\frac{\theta}{2}\right)
\;\;\;\;\;\;\;\;{\rm for}\;\;\; 2/3\,\pi< \theta \le
\pi\;.\label{lar} \ee A plot of the quantities $V$, $D$ and
$D^2+V^2$ is shown in the Figure. We see that something unexpected
happens: while in the interval $0 \le \theta < \pi/2$, $V$
decreases and $D$ increases, as expected from the wave-particle
duality, we see that in the interval $\pi/2 \le \theta \le \pi$,
$V$ and $D$ decrease and increase simultaneously! If we pick two
values $\theta_1$ and $\theta_2$ in this region, we obtain two
which-way detectors, that precisely realize the situation
described at the end of the Introduction.   It can also be seen
from the Figure (the dashed line) that the sum $D^2+V^2$ is
significantly less than one, for most values of $\theta$. We have
checked that these problems persist if, rather than $D$, one uses
the alternative  measure of which-way information $I_D$, provided
by Eq.(6.16) of Ref. \cite{durr}, since it turns out that the
optimal  observable for $I_D$ coincides with that relative to $D$,
in the interval $0 \le \theta < \,2/3\,\pi$.

In the  literature on the Quantum Detection problem, it has been
argued that it is sometimes possible to achieve a larger amount of
information on an unknown quantum state, by including an auxiliary
quantum system, called ancilla, in the read-out apparatus of a
quantum detector \cite{helstrom, peres}. This question has a
negative answer in the example above, but we do not touch upon
this problem here, and we refer the interested reader to
Ref.\cite{bimonte} for details.

\section{Discussion.}

In conclusion, the inequalities discovered by D\"urr in his
analysis of multibeam interferometers, are very interesting,
because they represent a set of testable relations between
measurable quantities, that follow directly from the first
principles of Quantum Mechanics. However, there is an important
difference between the two-beam relation, Eq.(\ref{D+V}), and its
multibeam generalization, Eq.(\ref{durr}). As we pointed out
above, the two-beam relation becomes an equality whenever the
beams and the detector are prepared in pure states, and this
entails the existence of a see-saw   relation between ${\cal D}$
and ${\cal V}$. We think that this behavior expresses the
intuitive idea of wave-particle duality, according to which
"...the more clearly we wish to observe the wave nature ...the
more information we must give up about... particle properties"
\cite{zurek}.  In other words Eq.(\ref{D+V}) conveys the basic
idea of interferometric duality, for which, in an ideal
interference experiment (namely one involving pure states) in
Quantum Mechanics,   ${\cal D}$ and ${\cal V}$ exhibit a dual
behavior. Any departure from this behavior, occurring  for mixed
states in the two-beam case, may be attributed to the presence of
extra sources of uncertainty, in addition to the unavoidable one
entailed by Quantum Mechanics.

In contrast, the inequality Eq.(\ref{durr}) is almost never
saturated, even for pure states \cite{bimonte}. So, while
Eq.(\ref{durr}) sets an upper bound for either quantity, when the
other takes a fixed value, it is not strong enough to prevent the
behavior exhibited in the  example presented in the previous
Section. According to it, even in an ideal experiment with pure
states, one can easily have cases when $D$ and $V$ both increase
or decrease at the same time. In the light of this, it seems to us
difficult to regard Eq.(\ref{durr}), as a statement of
interferometric duality, similar to  Englert's inequality for the
two-beam case. It is our opinion that the issue of giving a
complete Quantum Mechanical formulation of the interferometric
duality in multibeam experiments deserves further analysis.

\end{document}